\documentstyle[aps,twocolumn,epsf]{revtex}
\makeatletter
\@addtoreset{equation}{section}
\makeatother

\newcommand{\APP}{{Acta.\ Phys.\ Pol.\ }}

\newcommand{\EPJ}{{Eur.\ Phys.\ J.\ }}
\newcommand{\JMP}{{J.\ Math.\ Phys.\ }}
\newcommand{\NP}{{Nucl.\ Phys.\ }}
\newcommand{\PL}{{Phys.\ Lett.\ }}
\newcommand{\PR}{{Phys.\ Rev.\ }}
\newcommand{\PRep}{{Phys.\ Rep.\ }}
\newcommand{\PRL}{{Phys.\ Rev.\ Lett.\ }}
\newcommand{\RMP}{{Rev.\ Mod.\ Phys.\ }}

\def\CB {C\llap{/\kern3pt}}
\def\OB {O\llap{/\kern3pt}}
\def\PB {P\llap{/\kern1pt}}
\def\KB {K\llap{/\kern3pt}}
\def\pB {p\llap{/\kern1pt}}
\def\kB {k\llap{/\kern1pt}}
\def\qB {q\llap{/\kern1pt}}
\def\KB {K\llap{/\kern1pt}}
\def\FB {F\llap{/\kern1pt}}
\def\lB {l\llap{/\kern1pt}}
\def\LB {L\llap{/\kern1pt}}
\def\calDB {{\cal D}l\llap{/\kern1pt}}
\def\SigmaB {\Sigma\llap{/\kern1pt}}
\def\OmegaB {\Omega\llap{/\kern1pt}}
\def\barOmegaB {\bar\Omega\llap{/\kern1pt}}
\begin{document}
\thispagestyle{empty}
\vspace*{11mm}
\begin{center}
{\Large \bf Cancellation of pinching
singularities in out-of-equilibrium thermal field theory}\\
\vspace*{11mm}
{\large
 I.~Dadi\'c$^{1,2}$ }
\footnote{Talk given at the 5th International Workshop
 "Thermal Field Theories and their Applications",
August 10-14, 1998,
Regensburg, Germany}
\\[24pt]
$^1$ Ruder Bo\v{s}kovi\'{c} Institute,  Zagreb, Croatia\\
$^2$ Fakult\"at f\"ur Physik, Universit\"at Bielefeld , Germany \\
\vspace*{11mm}
\end{center}
\date{\today}
\begin{abstract}
Long standing problem in out-of-equilibrium thermal field
theories are pinching singularities. We find that the
expressions suspect of pinching require loop particles
to be on the mass shell. This fact, with the help of threshold
effect and similar effect due to spin, leads to the
elimination of pinching in single self-energy insertion
approximation to propagator in all propagators appearing in
QED and QCD under very mild restrictions on particle densities.

This, together with the cancellation of collinear singularities,
allows the extraction of useful physical information contained
in the imaginary parts of the two loop diagrams.

In some cases of interest ($\pi-\rho $ interaction, electro-weak interaction,
decay of Higgs particle, ...) none of the mentioned mechanisms
works and one has to resort to the resummed Schwinger-Dyson series.
These cases are more sensitive to the limitations related to the finite
time range.
\end{abstract}
\section{ Introduction}
Out of equilibrium thermal field theories have
recently attracted much interest. From the experimental point of
view, various aspects of heavy-ion collisions and the related hot
QCD plasma are of considerable interest, in particular the
supposedly gluon-dominated stage.

Contrary to the equilibrium case$^{(\cite{landsman,mleb})}$ where
pinch, collinear, and infrared problems
have been successfully controlled,
out of equilibrium
theory$^{(\cite{schwinger,keldysh,rammer})}$ has
suffered from them to these days. However, progress
has been made in this field, too.

Weldon$^{\cite{weldon11}}$ has observed that the out of equilibrium
pinch singularity does not cancel; hence it  spoils analyticity
and causality. The problem gets
worse with more than one self-energy insertions.

Bedaque has argued that in out of equilibrium theory
the time extension should be finite. Thus, the time integration
limits from $-\infty $ to $+\infty $, which are responsible
for the appearance of pinches, have to be abandoned as
unphysical$^{\cite{bedaque}}$. Similar argument, referring to the
Fermi's "golden rule" is given by Greiner and Leupold$^{\cite{gl}}$.

Le Bellac and Mabilat$^{\cite{lebellac}}$ have shown that pinching
singularity gives a contribution of order $g^2\delta n$, where
$\delta n$ is a deviation from equilibrium. They have
also found that collinear singularities cancel in scalar theory,
and in QCD using physical gauges$^{\cite{rl}}$, but not in the case of covariant
gauges.
Ni\'egawa$^{\cite{niegawacom}}$ has found that the pinch-like term
contains a divergent part that cancels collinear singularities
in the covariant gauge.

Altherr and Seibert have found that in massive $g^2\phi^3$ theory
pinch singularity does not occur owing to the kinematical
constraint$^{\cite{as}}$.

Altherr has suggested a regularization
method in which the propagator is modified
by the width $\gamma $ which is an arbitrary function of momentum
to be calculated in a self-consistent way. In $g^2\phi^4$ theory,
for small deviations from
equilibrium, $\gamma$ was found to be just the usual equilibrium
damping rate$^{\cite{altherr}}$.

This recipe has been justified in the resummed Schwinger-Dyson
series in various problems with
pinching$^{\cite{bdr,bdrs,bdrk,carrington,niegawa}}$.

Baier, Dirks, and Redlich$^{\cite{bdr}}$ have calculated the
$\pi-\rho $ self-energy contribution to
 the pion propagator,
regulating pinch contributions by the damping rate. In
subsequent papers with
Schiff$^{\cite{bdrs,bdrk}}$ they have calculated the quark propagator
within the HTL approximation$^{\cite{p,ebp,ft}}$; in the resummed
Schwinger-Dyson series, the pinch is naturally regulated by
$Im\Sigma_R$.

Carrington, Defu, and Thoma$^{\cite{carrington}}$ have found that
no pinch singularities appear in the
HTL approximation to the resummed photon propagator .

Ni\'egawa$^{\cite{niegawa}}$ has introduced the notion of renormalized
particle-number density. He has found that, in the appropriately
redefined calculation scheme, the amplitudes and reaction rates are
free from pinch singularities.

By pinching singularity we understand the contour passing
between two infinitely close poles:
\begin{equation}\label{pinch}
\int {dx\over (x+i\epsilon)(x-i\epsilon)}.
\end{equation}
When $\epsilon $ tends to zero, the integration path is
"pinched" between the two poles, and the expression is ill-defined.
Integration gives an $\epsilon^{-1} $ contribution  plus
regular terms. Decomposition
of $(x\pm i\epsilon)^{-1}$ into
$PP(1/x)\mp i\pi\delta (x)$, gives the related ill-defined
$\delta^2$ expression.

The following expression, which
is similar to (\ref{pinch}), corresponds
to the resummed Schwinger-Dyson series:
\begin{equation}\label{pinchlikesd}
\int dx{\omega(x)\over (x-\Sigma_R(x)+i\epsilon)
(x-\Sigma^*_R(x)-i\epsilon)},
\end{equation}
where $\omega(x)$ and $\bar \omega(x)$ (which
appears in (\ref{pinchlike1})) are, respectively, proportional to
$\Omega(x)$ and $\bar \Omega(x)$
where $\Omega(x)$, $\Sigma_R(x)$,  and $\bar \Omega(x)$
are  the components
of the self-energy matrix.

In expression (\ref{pinchlikesd}), pinching is
absent$^{\cite{bdr,bdrs,bdrk,carrington,niegawa}}$ if
$Im\Sigma_R(x_o)\neq0$
at a value of $x_o$
satisfying $x_o-Re\Sigma_R(x_o)=0$.

The expression corresponding to the single self-energy
insertion approximation to the propagator is similar to
(\ref{pinchlikesd}):
\begin{equation}\label{pinchlike1}
\int dx{\bar \omega(x)\over (x+i\epsilon)(x-i\epsilon)}.
\end{equation}
One can rewrite the integral as
\begin{equation}\label{rpinchlike1}
\int {dx\over 2}\biggl({1\over x+i\epsilon}
+{1\over x-i\epsilon}\biggr)
{\bar \omega(x)\over x}.
\end{equation}
If it happens that
\begin{equation}\label{rpinchlike2}
\lim_{x \rightarrow 0} {\bar \omega(x)\over x}=K<\infty ,
\end{equation}
then  the integral (\ref{rpinchlike1}) decomposes into two pieces
that, although possibly divergent, do not suffer from pinching.

There are two cases in which the function $\bar \omega(x)$ is even
identically zero in the vicinity of the $x=0$ point:
in thermal equilibrium, because of detailed balance
relations; in massive $g^2 \phi^3$ theory out of equilibrium,
owing to the mass shell condition$^{\cite{as}}$. The latter mechanism
also works in
out of equilibrium QED if a small photon mass $m_{\gamma} $
is introduced. However, this
elimination of pinching can be misleading: the domain of $x$, where
$\bar \omega(x)=0 $, shrinks to a point as
$m_{\gamma} \rightarrow 0$.
We shall show that the elimination of pinching also occurs
in the $m_{\gamma}=0$ case.

In this paper$^{\cite{id}}$ we identify two mechanisms leading to
relation (\ref{rpinchlike2}). They are based on the observation that
in the pinch-like contribution loop particles have to be on mass
shell.

The first mechanism is effective in out of equilibrium QED:
in the pinch-like contribution to the
electron propagator, phase space vanishes linearly
as $x \rightarrow 0$ . In the pinch-like contribution to the photon
propagator, the
domain of integration is shifted to infinity as $x \rightarrow 0$.
For distributions disappearing rapidly enough at large energies, the
contribution again vanishes linearly in the $x \rightarrow 0$ limit.
This mechanism is also valid in
QCD in the cases with massive quarks.

In out of equilibrium massless QCD, phase space
does not vanish, but
there is an alternative mechanism: the spinor/tensor structure
in all cases leads to relation (\ref{rpinchlike2}).

In a few cases, none of the mentioned mechanisms works and one has
to sum the Schwinger-Dyson series.
This is  the case of the $\pi-\rho $ loop in the
$\pi $ self-energy . Even in the limit of zero pion mass,
$\bar \omega(x)$ vanishes only as $|x|^{1/2}$ and relation
(\ref{rpinchlike2}) is not fulfilled. A similar problem appears in
electroweak interactions involving
decays of $Z$ and $W$ bosons, decay of Higgs particles, etc.
Another important case is massless $g^2\phi^3$ theory. In contrast to
massless QCD, massless $g^2\phi^3$ theory contains
no spin factor to provide a $q^2$ factor
necessary to obtain (\ref{rpinchlike2}).

The densities are restricted only mildly: they should be cut off
at high energies,
at least as $|k_o|^{-3-\delta}$, in order to obtain a finite total particle
density; for nonzero $k_o$, they should be finite;
for $k_o$ near zero, they should not diverge more rapidly
than $|k_o|^{-1}$, the electron (positron) distribution
should have a finite derivative.

Furthermore, we were unable to eliminate pinches
related to the double, triple, etc.,
self-energy insertion contributions to the propagator.

The resummed Schwinger-Dyson series must be free from pinching
as "for any system which moves towards thermal equilibrium and
thus behaves dissipatively, the full propagator must have some
finite width"$^{\cite{gl}}$.
\section{ Propagators and the Schwinger-Dyson equation}
We start$^{\cite{cshy,niemi}}$ by defining out of equilibrium
thermal propagators for bosons, in the case
when we can ignore the variations of slow variables in Wigner
functions$^{\cite{lebellac,bio}}$:
\begin{eqnarray}\label{D11}
&&D_{11}(k)=D^*_{22}(k)\cr
\nonumber\\
&&={i \over k^2-m^2+2i\epsilon|k_o|}+
2\pi \sinh^2\theta\delta(k^2-m^2),
\end{eqnarray}
\begin{eqnarray}\label{D12}
&&D_{12}(k)=2\pi \delta(k^2-m^2)\cr
\nonumber\\
&&
(\cosh^2 \theta \Theta(k_o)+\sinh^2 \theta \Theta(-k_o)),
\end{eqnarray}
\begin{eqnarray}\label{D21}
&&D_{21}(k)=2\pi \delta(k^2-m^2)\cr
\nonumber\\
&&
(\cosh^2 \theta \Theta(-k_o)+\sinh^2 \theta \Theta(k_o)).
\end{eqnarray}
The propagator  satisfies the important condition
\begin{equation}\label{sumD}
0=D_{11}-D_{12}-D_{21}+D_{22}.
\end{equation}
To obtain the corresponding relations for fermions, we only need
to make the substitution
\begin{equation}\label{b-f}
\sinh^2\theta(k_o) \rightarrow -\sin^2\bar\theta(k_o).
\end{equation}
In the case of equilibrium, we have
\begin{equation}\label{eqB}
\sinh^2\theta(k_o)=n_B(k_o)={1 \over \exp\beta |k_o|-1},
\end{equation}
and similarly for fermions.
Out of equilibrium, $n_B(k_o)$ and $n_F(k_o)$ will be some given
functions of $k_o$.

We transform to the Keldysh components
\begin{equation}\label{DR}
D_R(k)=-D_{11}+D_{21}={-i \over k^2-m^2+2i\epsilon k_o},
\end{equation}
\begin{equation}\label{DA}
D_A(k)=-D_R^*(k)=D_R(-k),
\end{equation}
\begin{eqnarray}\label{barD}
&&D_K(k)=D_{11}+D_{22}=h(k_o)(D_R-D_A)\cr
\nonumber\\
&&=2\pi \delta(k^2-m^2)(1+2\sinh^2\theta),\cr
\nonumber\\
&&
h(k_o)=-\epsilon(k_o)(1+2\sinh^2\theta).
\end{eqnarray}
Again for fermions
\begin{eqnarray}\label{barDF}
D_K(k)=2\pi \delta(k^2-m^2)(1-2\sin^2\bar\theta).
\end{eqnarray}
The proper self-energy satisfies the condition
\begin{equation}\label{sumSigma}
0=\Sigma_{11}+\Sigma_{12}+\Sigma_{21}+\Sigma_{22}.
\end{equation}
It is also transformed into the Keldysh form:
\begin{equation}\label{SigmaR}
\Sigma_R=-(\Sigma_{11}+\Sigma_{21}),~\Sigma_A=\Sigma_R^*,
\end{equation}
\begin{equation}\label{Omega}
\Omega=\Sigma_{11}+\Sigma_{22}.
\end{equation}
The "cutting rules" (refs.\cite {weldoncr,ksem}, see also ref.\cite{gelis}
for application of the rules out of equilibrium)
will convince us
that only on-shell loop-particle momenta contribute to
$Im\Sigma_R$ and $\Omega $.

The Schwinger-Dyson equation
\begin{equation}\label{Schwinger-Dyson}
{\cal G}=G+iG \Sigma {\cal G},
\end{equation}
can be written in terms of Keldysh components as
\begin{equation}\label{KeldishR}
{\cal G}_R=G_R+iG_R \Sigma_R {\cal G}_R,
\end{equation}
\begin{eqnarray}\label{Keldysh}
&&{\cal G}_K=G_K\cr
\nonumber\\
&&+i\left(G_A\Omega {\cal G}_R+G_K\Sigma_R{\cal G}_R +
		G_A\Sigma_A{\cal G}_K\right).
\end{eqnarray}

By expanding (\ref{KeldishR}), we deduce the contribution from the
single self-energy insertion to be of the form
\begin{equation}\label{psol2GRA}
{\cal G}_R\approx G_R+iG_R\Sigma_RG_R,
\end{equation}
which is evidently well defined, and the Keldysh component
suspected for pinching:
\begin{eqnarray}\label{psol2barG}
&&{\cal G}_K\approx G_K\cr
\nonumber\\
&&+iG_A\Omega G_R+iG_K\Sigma_RG_R
+iG_A\Sigma_AG_K.
\end{eqnarray}
The equation for ${\cal G}_R$
is simple and the solution is straightforward:
\begin{equation}\label{sol1GR}
{\cal G}_R={1 \over G_R^{-1}-i\Sigma_R}=-{\cal G}_A^*.
\end{equation}
To calculate ${\cal G}_K$, we can use the solution (\ref{sol1GR}):
\begin{equation}\label{sol1barG}
{\cal G}_K={\cal G}_A\left(h(q_o)(G_A^{-1}-G_R^{-1})
+i\Omega\right){\cal G}_R.
\end{equation}
The first term in (\ref{sol1barG}) is not always zero, but it
does not contain pinching singularities!
The second term in (\ref{sol1barG}) is potentially ill-defined (or
pinch-like).
The pinch-like contribution appears only in this equation;
thus it is the key to the
whole problem of pinch singularities. In the one-loop
approximation, it
requires loop particles to be on mass shell.

We start with (\ref{psol2barG}).
After substituting (\ref{barD})
into (\ref{psol2barG}),
we obtain the regular term plus the pinch-like contribution:
\begin{equation}\label{pert0}
{\cal G}_K\approx {\cal G}_{Kr}+{\cal G}_{Kp},
\end{equation}
\begin{eqnarray}\label{pinch0r}
&&{\cal G}_{Kr}=h(q_o)\cr
\nonumber\\
&&\left(G_R-G_A+iG_R\Sigma_RG_R
-iG_A\Sigma_AG_A\right),
\end{eqnarray}
\begin{equation}\label{pinch0p}
{\cal G}_{Kp}= iG_A\bar \Omega G_R,~~~
\bar \Omega=\Omega -h(q_o)(\Sigma_R-\Sigma_A).
\end{equation}
For equilibrium densities, we have
$\Sigma_{21}=e^{-\beta q_o}\Sigma_{12}$ , and
expression (\ref{pinch0p}) vanishes identically.

Expression (\ref{pinch0p}) is the only one suspected of pinch
singularities at the single self-energy insertion level.
The function $\bar\Omega $ in (\ref{pinch0p}) belongs to the type
of functions characterized by the fact that both loop particles
have to be on mass shell. It is analyzed in detail in Secs. III
and IV (for threshold effect) and in Sec. V (for spin effect).
With the help of this analysis we show that
relation (\ref{pinch0p}) transforms into
\begin{eqnarray}\label{pinch0pd}
&&{\cal G}_{Kp}= -i{K(q^2,q_o)\over 2}
\cr
\nonumber\\
&&\left({1\over q^2-m^2+2i\epsilon q_o}
+{1\over q^2-m^2-2i\epsilon q_o}\right),
\end{eqnarray}
where $K(q^2,q_o)$ is $\bar\Omega/(q^2-m^2)$ multiplied by
spinor/tensor factors included in the definition of $G_{R,A}$.
The finiteness of the limit
\begin{equation}\label{pinch0pdk}
\lim_{q^2\rightarrow m^2\mp 0}K(q^2,q_o) = K_{\mp}(q_o) < \infty,
\end{equation}
is important for cancellation of pinches. The index $\mp $ indicates
that the limiting value $m^2$ is
approached from either below or above, and these two values  are
generally different.
To isolate the potentially divergent terms, we express the function
$K(q^2,q_o)$ in terms of functions that are
symmetric ($K_{1}(q^2,q_o)$)
and antisymmetric ($K_2(q^2,q_o)$) around the value $q^2=m^2$:
\begin{eqnarray}\label{Kby12}
&&K(q^2,q_o)
=K_1(q^2,q_o)\cr
\nonumber\\
&&+\epsilon(q^2-m^2)K_2(q^2,q_o).
\end{eqnarray}
These functions are given by
\begin{eqnarray}\label{K12}
&&K_{1,2}(q^2,q_o)\cr
\nonumber\\
&&={1\over 2}\big(K(q^2,q_o)\pm K(2m^2-q^2,q_o)\big).
\end{eqnarray}
Locally (around the value $q^2=m^2$),
these functions are related to the limits $K_{\pm}(q_o)$ by
\begin{equation}\label{K12pm}
K_{1,2}(q^2,q_o)={1\over 2}\big(K_{+}(q_o)\pm K_{-}(q_o)\big).
\end{equation}
As a consequence, the right-hand side of
expression (\ref{pinch0pd}) behaves locally as
\begin{eqnarray}\label{pinchs}
&&{\cal G}_{Kp}(q^2,q_o)
\cr
\nonumber\\
&&\approx-{i\over 2}\left(K_1(q_o)+
\epsilon(q^2-m^2)K_2(q_o)\right)
\cr
\nonumber\\
&&\left({1\over q^2-m^2+2i\epsilon q_o}
+{1\over q^2-m^2-2i\epsilon q_o}\right),
\end{eqnarray}
and the term proportional to $K_2$ is capable of producing
logarithmic singularity.

\section{Threshold factor}
In this section we analyze the phase space of the loop integral
with both loop particles on mass shell. Special care is devoted
to the behavior of this integral near thresholds.
The expressions are written for all particles being bosons, and
spins are not specified; change to fermions is elementary.

Now, starting from (\ref{SigmaR}) to (\ref{Omega}), we calculate
$\Omega $ and $Im\Sigma_R$:
\begin{eqnarray}\label{Omegai2}
\Omega=2iIm\Sigma_{11}
=2{ig^2 \over 2}\int d\mu N_{\Omega }(k_o,k_o-q_o)F,
\end{eqnarray}
where
\begin{eqnarray}\label{dmu}
d\mu= {d^4k \over (2\pi)^4} 4\pi^2
\delta(k^2-m_D^2)\delta((k-q)^2-m_S^2),
\end{eqnarray}
and
\begin{eqnarray}\label{nomega}
&&N_{\Omega}(k_o,k_o-q_o)=
-.5\epsilon(k_o(k_o-q_o))\cr
\nonumber\\
& &+(.5+\sinh^2\theta_D(k_o))(.5+\sinh^2\theta_S(k_o-q_o)),
\end{eqnarray}
\begin{eqnarray}\label{SigmaRi2}
&&Im\Sigma_R={g^2 \over 2}\int d\mu
N_R(k_o,k_o-q_o)F,
\end{eqnarray}
and
\begin{eqnarray}\label{nr}
& &N_R(k_o,k_o-q_o)\cr
\nonumber\\
& &=\sinh^2\theta_D(k_o)\epsilon(k_o-q_o)
\cr
\nonumber\\
&&+\sinh^2\theta_S(k_o-q_o)\epsilon(-k_o)\cr
\nonumber\\
& &+\Theta(-k_o)\Theta(k_o-q_o)-\Theta(k_o)\Theta(q_o-k_o).
\end{eqnarray}
$F$ is the factor dependent on spin and
internal degrees of freedom.

It is useful to define $N_{\bar \Omega}(k_o,k_o-q_o)$ as
\begin{equation}\label{nbaromega}
N_{\bar \Omega}=
N_{\Omega}-h(q_o)N_R.
\end{equation}
After integrating over $\delta$'s, one obtains
\begin{eqnarray}\label{dmu1}
d\mu={1 \over 4|\vec q|}
{|k_o|dk_o \over |\vec k|}d\phi\Theta(1-z_o^2),
\end{eqnarray}
and expressions for $\Omega $ and $Im\Sigma_R$ take
general form
\begin{eqnarray}\label{trint1}
{\cal I}=\int d\mu N(k_o,k_o-q_o)
F(q,k_o,|\vec k|,\vec q\vec k),
\end{eqnarray}
where $|\vec k|=(k_o^2-m_D^2)^{1/2}$,
\begin{equation}\label{trint2}
\vec q\vec k=|\vec q||\vec k|z_o,
\end{equation}
\begin{equation}\label{trint3}
z_o={\vec q^2+\vec k^2-(\vec q-\vec k)^2 \over 2|\vec k||\vec q|}.
\end{equation}
$\phi\epsilon (0,2\pi)$ is the
angle between vector $\vec k_T$ and $x$ axes.

Let us start with the $q^2>0$ case.
Solution of $\Theta(1-z_o^2)$ gives the integration limits
\begin{eqnarray}\label{ko12}
&&k_{o1,2}={1 \over 2q^2}(q_o(q^2+m_D^2-m_S^2)\cr
\nonumber\\
&&\mp{1 \over 2q^2}
|\vec q|((q^2-q^2_{+tr})(q^2-q^2_{-tr}))^{1/2},
\end{eqnarray}
\begin{equation}\label{qmp}
q_{\pm tr}=|m_D \pm m_S|.
\end{equation}
Assume now that $q_{tr}\neq 0$. In this case, at the
threshold, the limits shrink to the value
\begin{eqnarray}\label{koko}
k_{o~tr}={q_o(q^2_{tr}+m_D^2-m_S^2) \over 2q^2_{tr}}.
\end{eqnarray}

We define the coefficient $c_1$ by
\begin{eqnarray}\label{c1}
c_1={1 \over 4|\vec q|}\int d\phi N(k_{otr},k_{otr}-q_{o})F.
\end{eqnarray}
Now the expression (\ref{trint1}) can be approximated by
\begin{eqnarray}\label{2}
& &{\cal I}\approx c_1(|\vec k|_{2}-|\vec k|_{1})\cr
\nonumber\\
& &\approx c_1(\Theta(q^2-q_{+tr}^2)+ \Theta(-q^2+q_{-tr}^2))\cr
\nonumber\\
& &
{q_o((q^2-q^2_{+tr})(q^2-q^2_{-tr}))^{1/2} \over q^2}.
\end{eqnarray}
Relation (\ref{2}) is the key to further discussion of the
threshold effect.

We obtain this also for higher dimension (D=6, for example).

Owing to (\ref{qmp}) and (\ref{2}), the function
${\cal I}(q^2,m_D^2,m_S^2)$
has the following properties
important for cancellation of pinches.

It vanishes between the thresholds, i.e., the domain
$(m_D-m_S)^2<q^2<(m_D+m_S)^2$ is forbidden (${\cal I}=0$). If it happens
that the bare mass $m^2$ belongs to this domain, the single self-energy
insertion will be free of pinching. In this case, multiple
(double, triple, etc.) self-energy insertions will
 also be free of pinching. Massive $\lambda\phi^3$ theory$^{\cite{as}}$
 is a good example of this case.

It is (in principle) different from zero in the allowed domain
$q^2<(m_D-m_S)^2$ and $(m_D+m_S)^2<q^2$. In this case, one cannot get rid
of pinching.
 This situation appears in the $\pi-\rho $
interaction$^{\cite{bdr}}$.

The behavior at the boundaries (i.e., in the allowed region near
the threshold) depends on the masses $m_D$ and $m_S$ and there are
a few possibilities.

If both masses are nonzero and different
($0\neq m_D\neq m_S\neq 0$), then there are two
thresholds and ${\cal I}$ behaves as $(q^2-q_{\pm tr}^2)^{1/2}$
in the allowed region near the threshold $q_{\pm tr}^2$. For
$m^2=q_{tr}^2$, the power
$1/2$ is not large enough to suppress pinching.

If one of the masses is zero ($m_D\neq0, m_S=0$ or
$m_D=0, m_S\neq 0$), then (\ref{2}) gives that the thresholds are
identical (i.e., the forbidden domain shrinks to zero) and one obtains
the $(q^2-m_D^2)^1$ behavior near $m_D^2$. This case (for $m^2=m_D^2$)
is promising. The elimination of pinching in the electron propagator,
considered in Sec.IV, is one of important examples.

If the masses are equal but different from zero
($m_D=m_S\neq 0$), then there are two
thresholds with different behavior. The function ${\cal I}$
behaves as $(q^2-q_{+tr}^2)^{1/2}$
in the allowed region near the threshold $q_{+tr}^2=4m_D^2$,and this
behavior cannot eliminate pinching in the supposed case $m^2=4m_D^2$ .

However, at the other threshold, namely at
 $q^2_{-tr}=0$, the physical region is determined
by $q^2<0$ and the above discussion does not apply.
In fact, the integration limits (\ref{ko12}) are
valid, but the region between $k_{o~1}$ and $k_{o~2}$
is now excluded from integration.
One has to integrate  over the domain
$(-\infty,k_{o~1})\bigcup (k_{o~2},+\infty)$.
This leads to the limitation in the high-energy behavior of the
density functions. An important example of such behavior, elimination of
pinching in the photon propagator ($m_{\gamma}$), is discussed in Sec.IV.

If both masses vanish ($m_D=m_S=0$), the thresholds coincide,
there is no forbidden region and no threshold behavior. The
behavior depends on the spin of the particles involved.
For
scalars, the leading term in the expansion of ${\cal I}$ does
not vanish. Pinching is not eliminated.

The case of vanishing masses ($m_D=m_S=0$) for particles with spin
exhibits a peculiar behavior. In all studied examples
(see Sec.V for details),
 ${\cal I}$ behaves as $q^2$ as $q^2\rightarrow 0$, which promises
the elimination of pinching.

\section{ Pinch Singularities in QED}
\subsection{ Pinch Singularities in the Electron Propagator}
In  this subsection we apply the results of preceding section
to cancel the pinching singularity appearing in a single
self-energy insertion approximation to the electron propagator.
To do so, we have to substitute $m_D=m$, $m_S=0$,
$\sinh^2\Theta_D(k_o)\rightarrow -n_e(k_o)$,
$\sinh^2\Theta_S(k_o-q_o) \rightarrow n_\gamma(k_o-q_o)$,
and $h(k_o)=-\epsilon(k_o)(1-2n_e(k_o))$,
where $n_e$ and $n_\gamma$ are given non-equilibrium distributions of
electrons and photons in relations (\ref{nomega}),
(\ref{nr}),(\ref{nbaromega}), and (\ref{barD}).
The thresholds are now identical
($q^2_{\pm tr}=m^2$),
and the integration limits satisfy
\begin{equation}\label{ke12}
|\vec k|_{2}-|\vec k|_1={q_o \over q^2}(q^2-m^2)).
\end{equation}
At threshold the limits shrink to the value
$k_{o~tr}=q_o,~|\vec k|_{tr}=|\vec q|$.

Then, with the help of (\ref{c1}), we define
\begin{eqnarray}\label{KB}
& &\KB(q^2,q_o)={(\qB+m)\barOmegaB(\qB+m)\over (q^2-m^2)}\cr
\nonumber\\
& &\approx
{1 \over 16\pi^2|\vec q|(q^2-m^2)}\int d\phi N_{\bar \Omega}
(k_{otr},k_{otr}-q_o)
\cr
\nonumber\\
& &
(\qB+m)\FB(\qB+m)(|\vec k|_{2}-|\vec k|_{1}).
\end{eqnarray}

For $q^2\neq 0$, we can decompose the vector $k$ as
\begin{eqnarray}\label{kl}
&&k={(k.q)\over q^2}q+{(k.\tilde q)\over \tilde q^2}\tilde q+k_T
\cr
\nonumber\\
& &=(q-{q_o\over |\vec q|}\tilde q){-m_\gamma^2+m^2+q^2\over 2q^2}+
{k_o\over |\vec q|}\tilde q+k_T,
\end{eqnarray}

where, in the heat-bath frame  we have
\begin{eqnarray}\label{tildeq}
&&q=(q_o,0,0,|\vec q|),~~
\tilde q=(|\vec q|, 0,0,q_o),
\cr
\nonumber\\
& &q\tilde q=0,~~\tilde q^2=-q^2.
\end{eqnarray}
In calculating the term proportional to $(1-a)$, where $a$ is
the gauge parameter, we have to use
the trick
\begin{eqnarray}\label{mgamma}
&&((k-q)^2\pm i\epsilon)^{-2}
\cr
\nonumber\\
& &=\lim_{m_\gamma \rightarrow 0}\left[
{\partial \over \partial m_\gamma^2}
((k-q)^2\pm i\epsilon)-m_\gamma^2)^{-1}\right].
\end{eqnarray}
Finally, we obtain The "sandwiched"
trace factor $\FB $ calculated with loop particles on mass
shell:
\begin{eqnarray}\label{qfq}
& &(\qB+m)\FB(\qB+m)=2m(q^2+m^2+2m\qB)\cr
\nonumber\\
& &
+(q^2-m^2)\biggl(-{q^2-m^2\over q^2}\qB
\cr
\nonumber\\
& &+(-{q_o(q^2+m^2)\over q^2|\vec q|}+2{k_o\over |\vec q|})
\tilde \qB+2\kB_T
\cr
\nonumber\\
& &-(1-a){(q^2-m^2)\over 2q^2}(-\qB+{q_o\over |\vec q|}
\tilde \qB)\biggr).
\end{eqnarray}
Now we can study the limit
\begin{eqnarray}\label{KkB}
&&\KB(q_o)=\lim_{q^2\rightarrow m^2}\KB(q^2,m^2,q_o)
\cr
\nonumber\\
&&=
(\qB+m){q_o \over 2\pi|\vec q|m^2}N_{\bar \Omega}
(k_{o~tr},k_{o~tr}-q_{o}).
\end{eqnarray}
It is easy to find that $\KB(q_o) $ is finite provided that
$m^2\neq 0$ and
$N_{\bar \Omega}(q_o,0)<\infty $. The last condition
is easy to investigate using the limiting procedure:
\begin{eqnarray}\label{egamma}
& &N_{\bar \Omega}(q_o,0)=\lim_{k_o\rightarrow q_o}
N_{\bar \Omega}(k_o,k_o-q_o)
\cr
\nonumber\\
& &=\lim_{k_o\rightarrow q_o}2n_\gamma(k_o-q_o)(n_e(q_o)-n_e(k_o))
\cr
\nonumber\\
& &
+\lim_{k_o\rightarrow q_o}(n_e(q_o)-n_e(k_o)
-\epsilon(q_o)\epsilon(k_o-q_o)\cr
\nonumber\\
& &\lim_{k_o\rightarrow q_o}(n_e(q_o)+n_e(k_o)-2n_e(q_o)n_e(k_o)
.
\end{eqnarray}
The integration limits imply that
the limit $k_o\rightarrow q_o$ is taken from below for $q^2>m^2$,
and from above for $q^2<m^2$. The two limits lead to different
values of $N_{\bar \Omega}(q_o,0)$. This leads to the
discontinuity of
$\KB(q^2,m^2,q_o)$ at the point $q^2=m^2$.

Only the first term in (\ref{egamma}) can give rise to problems. We
rewrite it as $\lim_{k_o\rightarrow 0}\left(2k_on_\gamma(k_o)
{\partial n_e(k_o+q_o)\over \partial k_o}\right)$.
As relation
(\ref{KkB}) should be valid at any $q_o$ we find two conditions:
\begin{equation}\label{cgamma}
\lim_{k_o\rightarrow 0}k_on_\gamma(k_o)<\infty,
\end{equation}
\begin{equation}\label{cel}
|{\partial n_e(q_o)\over \partial q_o}|<\infty.
\end{equation}
Under the very reasonable conditions (\ref{cgamma}) and (\ref{cel})
the electron propagator is free from pinching.

It is worth observing that $\KB(q_o) $ is gauge independent, at
least within the class of covariant gauges.

\subsection{ Pinch Singularities in the Photon Propagator}
To consider the pinching singularity appearing in a single
self-energy insertion
approximation to the photon propagator, we have to make
the substitutions
$m_D=m=m_S$, $\sinh^2\Theta_D(k_o)\rightarrow -n_e(k_o)$,
$\sinh^2\Theta_S(k_o-q_o) \rightarrow -n_e(k_o-q_o)$,
and $h(k_o)=-\epsilon(k_o)(1+2n_\gamma(k_o))$.
There are two thresholds, but only $q^2_{1, tr}=0$ and the domain
where $q^2<0$ are relevant  to a
massless photon.
The integration limits are given by the same expression
(\ref{ko12}), but now we have to integrate over
the domain $(-\infty,k_{o~1})\bigcup (k_{o~2},+\infty)$.
As $q^2 \rightarrow -0$, we
 find $(k_{o~1}\rightarrow -\infty)$ and
$(k_{o~2}\rightarrow +\infty)$.
The integration domain is still infinite but is shifted toward
$\pm \infty$ where one expects that the particle
distribution vanishes:
\begin{eqnarray}\label{Kmn}
& &K_{\mu\nu}(q^2,q^o)=
\left(g_{\mu\rho}-(1-a)
{q_{\mu}q_{\rho}\over q^2-2iq_o\epsilon}\right )
\cr
\nonumber\\
& &{\bar\Omega^{\rho\sigma}\over q^2}
\left(g_{\sigma \nu}-(1-a)
{q_{\sigma}q_{\nu}\over q^2+2iq_o\epsilon}\right )
\cr
\nonumber\\
& &=
{1 \over 16\pi^2|\vec q|q^2}
\left(\int_{-\infty}^{k_{o1}}+\int_{k_{o2}}^{\infty}
\right){k_odk_o\over |\vec k|}
\int d\phi
\cr
\nonumber\\
& &N_{\bar \Omega}(k_o,k_o-q_o)
\left(g_{\mu\rho}-(1-a)
{q_{\mu}q_{\rho}\over q^2-2iq_o\epsilon}\right )
\cr
\nonumber\\
& &
F^{\rho\sigma}\left(g_{\sigma \nu}-(1-a)
{q_{\sigma}q_{\nu}\over q^2+2iq_o\epsilon}\right ).
\end{eqnarray}
To calculate $F^{\mu\nu}$ for the $e-\bar e$ loop, we
parameterize the loop momentum $k$ by introducing
an intermediary variable $l$ perpendicular to $q$.
$m$ is the mass of loop particles:
\begin{eqnarray}\label{lm}
&&k=\alpha q+l,~q.l=o,~
k^2=(k-q)^2=m^2,~
\cr
\nonumber\\
&&l^2=m^2-\alpha^2q^2,~\alpha={k^2+q^2-(k-q)^2 \over 2q^2}.
\end{eqnarray}
After all possible singular denominators are canceled,
one can set $\alpha=1/2$.
\begin{eqnarray}\label{fmn}
& &F_{e\bar e}^{\mu \nu}=
-Tr(\kB +m)\gamma^{\mu}(\kB-\qB+m)\gamma^{\nu}\cr
\nonumber\\
& &=\biggl({4m^2q_o^2\over \vec q^2}A^{\mu \nu}(q)\cr
\nonumber\\
& &
+{q^2 \over \vec q^2}\biggl((4k_o(k_o-q_o)-4m^2-q^2)A^{\mu \nu}(q)
\cr
\nonumber\\
& &+(-8(k_o-{q_o\over 2})^2+2\vec q^2) B^{\mu \nu}(q)\biggr)\biggr),
\end{eqnarray}
For projection operators $A$, $B$, $C$ and  $D$ see (\ref{A})-(\ref{Dmunu}).
Now we obtain
\begin{eqnarray}\label{Kmns}
& &K_{\mu\nu}(q^2,q_o)=
{1 \over 16\pi^2|\vec q|q^2}\cr
\nonumber\\
& &
\left(\int_{-\infty}^{k_{o1}}+\int_{k_{o2}}^{\infty} \right)
{k_odk_o\over |\vec k|}
\int d\phi N_{\bar \Omega}(k_o,k_o-q_o)
\cr
\nonumber\\
& &
\biggl({4m^2q_o^2\over \vec q^2}A_{\mu \nu}(q)\cr
\nonumber\\
& &
+{q^2 \over \vec q^2}\biggl((4k_o(k_o-q_o)-4m^2-q^2)A_{\mu \nu}(q)
\cr
\nonumber\\
& &+(-8(k_o-{q_o\over 2})^2+2\vec q^2) B_{\mu \nu}(q)\biggr)\biggr).
\end{eqnarray}
In the integration over $k_o$ the terms proportional to
$(k_o^2q^2)^n$ dominate and
$\lim_{q^2\rightarrow 0}|K_{\mu\nu}(q^2,q_o)|<\infty$
if
\begin{eqnarray}\label{Kmna}
& &|{1 \over 16\pi^2|\vec q|q^2}
\left(\int_{-\infty}^{k_{o1}}+\int_{k_{o2}}^{\infty} \right)
{k_odk_o\over |\vec k|}\cr
\nonumber\\
& &(\alpha+\beta k_o^2 q^2)
\int d\phi N_{\bar \Omega}(k_o,k_o-q_o)|<\infty.
\end{eqnarray}
Here $N_{\bar \Omega}(k_o,k_o-q_o)$ is given by
\begin{eqnarray}\label{ebare}
& &N_{\bar \Omega}(k_o,k_o-q_o)=
-2n_{e}(k_o-q_o)\cr
\nonumber\\
& &(-n_\gamma(q_o)-n_e(k_o))
-n_\gamma(q_o)-n_e(k_o)
\cr
\nonumber\\
& &-\epsilon(q_o)\epsilon(k_o-q_o)(-n_\gamma(q_o)
+n_e(k_o)
\cr
\nonumber\\
& &+2n_\gamma(q_o)n_e(k_o)).
\end{eqnarray}
Assuming that the distributions obey the inverse-power law at large
energies $n_{e}(k_o)\propto |k_o|^{-\delta_e}$ and
$n_{\bar e}(k_o)\propto |k_o|^{-\delta_{\bar e}}$, we find that
the terms linear in densities dominate. Thus, for $n=0,1$, one finds
\begin{eqnarray}\label{ltsa}
&&{-1\over q^2}\biggl(\int_{-\infty}^{k_{o~1}}
+\int_{k_{o~2}}^{+\infty}\biggr)
{|k_o|dk_o\over |\vec k|}|k_o|^{2n-\delta} (-q^2)^n
\cr
\nonumber\\
&&\propto (\delta-1-2n)^{-1}(|\vec q|m)^{1+2n-\delta}
(-q^2)^{(\delta-3)/2}.
\end{eqnarray}
It follows that (\ref{Kmna}) is finite (in fact, it vanishes)
if $\delta_e,\delta_{\bar e}>3$. Similar analysis for electron
propagator at $q^2 < 0$ (thus outside of our analysis of
pinch singularities) leads to $\delta_{\gamma}>3$.
This is exactly the condition
\begin{equation}\label{fine}
\int d^3kn_{\gamma,e,\bar e}(k_o)<\infty.
\end{equation}
Thus the pinching singularity is canceled in the photon propagator
under the condition that the electron and positron distributions
should be such that the total number of particles is finite.

Also, in the photon propagator,
the quantity $\lim_{q^2 \rightarrow 0}K_{\mu\nu}(q^2,q_o)$
does not depend on the gauge parameter.

Expression (\ref{ltsa}) is not valid for $m=0$.

\section{ Pinch Singularities in Massless QCD}
In this section we consider the case of massless QCD.
One should observe that the massless quarks and gluons are an
idealisation eventually appropriate at the lowest order.
In the nonequilibrium HTL resummation scheme both quarks and gluons
acquire dynamical mass$^{\cite{carrington}}$.
Pinching singularities, related to massive quarks,
are eliminated by the methods used in the preceding section.

Attention is turned to
the spin degrees of freedom, i.e., to the function $F$ of the
integrand in (\ref{Omegai2}) to (\ref{trint1}). In the calculation
of $F$ it has been anticipated that the loop particles have to be
on mass shell. In this case, $F$ provides an extra $q^2$ factor
which suffices for the elimination of pinching
singularities.

The integration limits are now
\begin{equation}\label{ko12qg}
k_{o1,2}={1\over 2}\left(q_o\mp|\vec q|\right).
\end{equation}
The difference $|\vec k|_2-|\vec k|_1$ is finite and
there is no threshold effect.

It is worth observing that for $q^2>0$, we have to integrate between
$k_{o1}$ and $k_{o2}$, whereas for $q^2<0$, the integration domain is
$(-\infty,k_{o~1})\bigcup (k_{o~2},+\infty)$. This leads to two
limits, $\lim_{q^2\rightarrow \pm0}K(q^2,q_o)=K_{\pm}(q_o)$,
in all cases of massless QCD.

By inspection of the final results (\ref{qqse}),(\ref{ghghse}), and
(\ref{ggse}),  we find that the case $q^2<0$ requires integrability
of the function $k_o^2N_{\bar\Omega}(k_o,k_o-q_o)$ leading to the
condition (\ref{fine})  on the quark, gluon, and ghost distribution
functions.

The function $K_{\mu \nu}(q^2,q_o)$ related to the gluon propagator
is the sum  of the contributions from various loops,
where the terms in the sum are defined as
\begin{eqnarray}\label{VCbarOS}
& &K_{\mu \nu}(q^2,q_o)=
(g_{\mu\rho}-(1-a)D_{R\mu\rho})\cr
\nonumber\\
& &
{\bar\Omega^{\rho \sigma} \over q^2}
(g_{\sigma\nu}-(1-a)D_{A\sigma\nu}).
\end{eqnarray}
The tensor $F$ related to the massless quark-antiquark
contribution to the gluon self-energy is
\begin{eqnarray}\label{qqse}
& &F_{q \bar q}^{\mu \nu}=
-{\delta_{a b} \over 6}Tr\kB\gamma^{\mu}(\kB-\qB)\gamma^{\nu}
\cr
\nonumber\\
& &={\delta_{a b} \over 6}
\biggl({q^2 \over \vec q^2}\biggl((4k_o(k_o-q_o)-q^2)A^{\mu \nu}(q)
\cr
\nonumber\\
& &+(-8(k_o-{q_o\over 2})^2+2\vec q^2) B^{\mu \nu}(q)\biggr)
+O^{\mu\nu}(\vec k_T)\biggr).
\end{eqnarray}
As $F_{\mu\nu}$ contains only $A$ and $B$ projectors,
the result does not depend on the gauge parameter.

Relation (\ref{qqse}) contains only terms proportional to $q^2$,
and $\lim_{q^2\rightarrow 0}K_{\mu\nu}(q^2,q_o)$ is finite.

For the ghost-ghost contribution to the gluon self-energy,
the tensor $F$ is given by
\begin{eqnarray}\label{ghghse}
& &F_{gh gh}^{\mu \nu}=-\delta_{a b}N_ck^{\mu}(k-q)^{\nu}
\cr
\nonumber\\
& &=
-\delta_{a b}N_c{q^2 \over \vec q^2}
\biggl({4k_o(k_o-q_o)+q^2 \over 8}A^{\mu \nu}(q)
\cr
\nonumber\\
& &-(k_o-{q_o \over 2})^2 B^{\mu \nu}(q)
-{\vec q^2 \over 4}D^{\mu \nu}(q)+O^{\mu\nu}(\vec k_T)
 \biggr).
\end{eqnarray}

The tensor $F$ for the gluon-gluon contribution to the gluon
self-energy is
\begin{eqnarray}\label{ggse}
& &F_{gg}^{\mu \nu}=
{\delta_{ab}N_c \over 2}\cr
\nonumber\\
& &\left(g^{\mu \sigma}(q+k)^{\tau}-
g^{\sigma \tau}(2k-q)^{\mu}+
g^{\tau \mu}(k-2q)^{\sigma}\right)\cr
\nonumber\\
& &
\left(g_{\sigma \rho}-(1-a){(k-q)_{\sigma}(k-q)_{\rho}
\over (k-q)^2\pm 2i(k_o-q_o)\epsilon}\right)\cr
\nonumber\\
& &\left(g^{\nu \rho}(q+k)^{\eta}-
g^{\rho\eta}(2k-q)^{\nu}+g^{\eta\nu}(k-2q)^{\rho}\right)
\cr\nonumber\\& &
\left(g_{\tau \eta}-(1-a){k_{\tau}k_{\eta}
\over k^2\pm 2ik_o\epsilon}\right)\cr
\nonumber\\
& &
\rightarrow{\delta_{ab}N_c q^2\over 2}\biggl({1\over \vec q^2}
\biggl((10(k_o-{q_o \over 2})^2+{3 \over 2}\vec q^2 )A^{\mu \nu}(q)
\cr
\nonumber\\
& &
+(-10(k_o-{q_o\over 2})^2+4\vec q^2) B^{\mu \nu}(q)
-{\vec q^2 \over 2}D^{\mu \nu}(q)\biggr)
\cr
\nonumber\\
& &
-(1-a)\biggl({1\over 2}A^{\mu\nu}-B^{\mu\nu}
-{q_o\over |\vec q|}C^{\mu\nu} \biggr)
\cr
\nonumber\\
& &
+(1-a)^2\biggl(-{q^2\over \vec q^2}A^{\mu\nu}
+2{q_o^2\over \vec q^2}B^{\mu\nu}
-2{q_o\over |\vec q|}C^{\mu\nu}\cr
\nonumber\\
& &-2D^{\mu\nu}\biggr)
+O^{\mu\nu}(\vec k_T)\biggr).
\end{eqnarray}
Expressions (\ref{qqse}), (\ref{ghghse}), and (\ref{ggse})
for the ghost-ghost, quark-antiquark, and gluon-gluon
contributions to the gluon self-energy contain only terms
proportional to $q^2$. The function $K_{\mu\nu}(q^2,q_o)$
approaches the finite value $K_{\mu\nu}(\pm,q_o)$.

Thus we have shown that the single
self-energy contribution to the gluon propagator is free from pinching
under the condition (\ref{fine}) .

The $K$ spinor for the quark-gluon contribution
to the massless quark propagator is defined as
\begin{equation}\label{SKbarOS}
\KB(q^2,q_o)=\qB{\barOmegaB\over q^2}\qB.
\end{equation}
In the self-energy of a massless quark coupled to a
gluon the "sandwiched" spin factor $\qB\FB\qB$ is given by
(as the term proportional to $\kB_T$ vanishes after integration, we drop it).
\begin{eqnarray}\label{qtrl}
&&\qB\FB_{qg}\qB=\delta_{a b}{N_c^2-1 \over 2N_c}
\cr
\nonumber\\
&&\left (g_{\mu \nu}-{(1-a)(k-q)_{\mu}(k-q)_{\nu}
\over (k-q)^2\pm 2i(k_o-q_o)\epsilon}\right )
\qB\gamma^\mu \kB\gamma^\nu\qB\cr
\nonumber\\
&&=\delta_{a b}{N_c^2-1 \over 2N_c}q^2
\cr
\nonumber\\
&&\left(-\qB-{q_o\over |\vec q|}\tilde\qB+2{k_o\over |\vec q|}
\tilde\qB
-{1-a\over 2}(-\qB+{q_o\over |\vec q|}\tilde\qB)\right),
\end{eqnarray}
which contains the damping factor $q^2$.

By inserting (\ref{qtrl}) into (\ref{SKbarOS}), we
obtain (\ref{pinch0pdk}) free from pinches.

To calculate $\KB(q_o)$,
we need the limit
\begin{equation}\label{qtrll}
\lim_{q^2\rightarrow 0}{\qB \FB_{qg}\qB\over q^2}
=\delta_{a b}{N_c^2-1 \over 2N_c}{2(k_o-q_o)\over q_o}\qB.
\end{equation}
From (\ref{qtrll}) we conclude that $\KB(q_o)$ does not depend
on the gauge parameter.

Omitting details, we observe that pinching is absent from
the quark propagator, also in the Coulomb gauge, with the same limit
(\ref{qtrll}).

The $K$ factor for the ghost-gluon self-energy
contribution to the ghost propagator is defined as
\begin{equation}\label{SgKbarOS}
K(q^2,q_o)={\bar\Omega\over q^2}.
\end{equation}
The $F$ factor for the ghost-gluon contribution  is
\begin{eqnarray}\label{ghgse}
&&F_{gh g}=\delta_{a b}N_ck^{\mu}q^{\nu}\cr
\nonumber\\
&&\left(g_{\mu \nu}-{(1-a)(k_{\mu}-q_{\mu})(k_{\nu}-q_{\nu})
\over (k-q)^2\pm 2i(k_o-q_o)\epsilon}\right)\cr
\nonumber\\
&&
\rightarrow \delta_{a b}N_c{q^2 \over 2}.
\end{eqnarray}
The factor $q^2$ ensures the absence of pinch singularity and a
well-defined perturbative result.

The $K$ factor for the scalar-photon self-energy
contribution to the scalar propagator is defined as
\begin{equation}\label{SbKbarOS}
K(q^2,q_o)={\bar\Omega\over q^2}.
\end{equation}
The $F$ factor for the massless scalar-photon contribution to the
scalar self-energy,
\begin{eqnarray}\label{scphot}
&&F_{s\gamma}=(q+k)^{\mu}(q+k)^{\nu}\cr
\nonumber\\
&&\left(g_{\mu \nu}-{(1-a)(k-q)_{\mu}(k-q)_{\nu} \over
(k-q)^2\pm 2i(k_o-q_o)\epsilon}\right)\rightarrow 2q^2,
\end{eqnarray}
clearly exhibits the $q^2$ damping factor!

\section{ Conclusion}
Studying the out of equilibrium Schwinger-Dyson equation, we
have found that ill-defined pinch-like expressions appear
exclusively in the Keldysh component (${\cal G}_K$) of the resummed
propagator (\ref{sol1barG}), or in the single self-energy insertion
approximation to it (\ref{pinch0p}).
This component does not vanish only in the expressions with
the Keldysh component (\ref{Omega}) ($\Omega $ or
$\bar\Omega $ for the single self-energy approximation)
of the self-energy matrix. This then requires that loop particles
be on mass shell. This is the crucial point to eliminate pinch
singularities.

We have identified two basic mechanisms for the elimination of
pinching: the threshold and the spin effects.

For a massive electron and a massless photon (or quark and gluon)
it is the threshold effect in the phase space integration
that produces, respectively, the critical $q^2-m^2$ or $q^2$
damping factors.

In the case of a massless quark, ghost, and gluon, this
mechanism fails, but the spinor/tensor structure of the self-energy
provides an extra $q^2$ damping factor.

We have found that, in QED, the pinching singularities appearing
in the single self-energy insertion approximation to the electron
and the photon propagators are absent under very reasonable
conditions: the distribution function should be finite,
exceptionally the photon distribution is allowed to diverge as
$k_o^{-1}$ as $k_o \rightarrow 0$; the
derivative of the electron distribution should be finite; the
total density of electrons should be finite.

For QCD, identical conditions are imposed on the distribution of
massive quarks and the distribution of gluons; the distributions of
massless quarks and ghosts (observe here that in the covariant
gauge, the ghost distribution is not required to be identically zero)
should be integrable functions;
they are limited  by the finiteness of the total density.

In the preceding sections we have shown that all pinch-like
expressions appearing in QED and QCD (with massless and massive
quarks!)  at the single self-energy insertion level
do transform into well-defined expressions. Many other theories
behave in such a way.
However, there are important exceptions:
all theories in which lowest-order processes are
kinematically allowed do not acquire well-defined expressions at
this level. These are electroweak interactions, processes
involving Higgs  and two light particles, a $\rho $ meson and two
$\pi $ mesons, $Z$, $W$, and other heavy particles decaying into
a pair of light particles, etc. The second important exception is
massless $g^2\phi^3$ theory. This theory, in contrast to
massless QCD, contains no spin factors to provide
(\ref{rpinchlike2}).
In these cases, one has to resort to the resummed Schwinger-Dyson
series. One can also expect that, in these cases, higher order
contributions become more important and provide natural cutoff
which reduces the contribution of pinch-like terms. In ultimate case
this points
 out to the limitations of the method.

The main result of the present paper is the cancellation of pinching
singularities at the single self-energy insertion level in QED- and
QCD-like theories. This, together with the
reported$^{\cite{lebellac,niegawacom}}$ cancellation of collinear
singularities, allows the extraction of useful physical
information contained in the imaginary parts of the two-loop diagrams.
 This is not the case with three-loop diagrams, because some of
them contain double self-energy insertions. In this case, one again
has to resort to the sophistication of resummed propagators.

\section{ Appendix}
We start$^{\cite{landsman}}$ by defining a heat-bath four-velocity
$U_{\mu}$, normalized to unity, and define the orthogonal projector
\begin{equation}\label{Delta}
\Delta_{\mu \nu}=g_{\mu \nu}-U_\mu U_\nu.
\end{equation}
We further define spacelike vectors in the heat-bath frame:
\begin{equation}\label{kappa}
\kappa_{\mu}=\Delta_{\mu \nu}q^\nu,~~~~\kappa_\mu \kappa^\mu
=\kappa^2=-\vec q^2.
\end{equation}

There are four independent symmetric tensors (we distinguish
retarded from advanced tensors by the usual
modification of the $i\epsilon $ prescription)
$A$, $B$, and $D$ (which are mutually orthogonal projectors), and $C$:
\begin{equation}\label{A}
A_{\mu \nu}(q)=
\Delta_{\mu \nu}-{\kappa_\mu \kappa_\nu \over \kappa^2},
\end{equation}
\begin{equation}\label{B}
B_{R~\mu \nu}(q)=U_\mu U_\nu +{\kappa_\mu \kappa_\nu
\over \kappa^2}-{q_\mu q_\nu \over (q^2+2iq_o\epsilon)},
\end{equation}
\begin{eqnarray}\label{Cmunu}
& &C_{R~\mu \nu}(q)
={(-\kappa^2)^{1/2} \over U.q}\cr
\nonumber\\
& &
\left ({(U.q)^2 \over \kappa^2}U_\mu U_\nu -{\kappa_\mu \kappa_\nu
 \over \kappa^2}+{(q_o^2+\vec q^2)
q_\mu q_\nu \over \vec q^2(q^2+2iq_o\epsilon)}\right ),
\end{eqnarray}
\begin{equation}\label{Dmunu}
D_{R~\mu \nu}(q)={q_\mu q_\nu \over q^2+2iq_o\epsilon}.
\end{equation}

In addition to the known multiplication$^{\cite{landsman}}$
properties (for convenience we drop $q$-dependence)
\begin{equation}\label{AA}
A A =A ,~B_{R,A} B_{R,A} =B_{R,A} ,
\end{equation}
\begin{eqnarray}\label{CRCR}
& &C_{R,A} C_{R,A} =-(B_{R,A} +D_{R,A} ),
\cr
\nonumber\\
& &D_{R,A} D_{R,A} =D_{R,A} ,
\end{eqnarray}
\begin{eqnarray}\label{AB}
& &A B =B A =A C =C A =0,
\cr
\nonumber\\
& &
A D =D A =B D =D B =0,
\end{eqnarray}
\begin{eqnarray}\label{BC}
& &(B_{R,A} C_{R,A} )_{\mu\nu}=(C_{R,A} D_{R,A} )_{\mu\nu}
\cr
\nonumber\\
& &
=(C_{R,A} B_{R,A} )_{\nu\mu}=(D_{R,A} C_{R,A} )_{\nu\mu}
\cr
\nonumber\\
& &
=
{\tilde q_\mu q_\nu\over q^2\pm 2iq_o\epsilon},
\end{eqnarray}
we need mixed products
\begin{equation}\label{BRBA}
B_{R,A} B_{A,R} ={1 \over 2}(B_R +B_A ),
\end{equation}
\begin{eqnarray}\label{CRCA}
C_{R,A} C_{A,R}=-{1 \over 2}(B_R +B_A+D_R +D_A ),
\end{eqnarray}
\begin{equation}\label{DRDA}
D_{R,A} D_{A,R} ={1 \over 2}(D_R +D_A ),
\end{equation}
\begin{eqnarray}\label{BCRA}
& &(B_{R,A} C_{A,R} )_{\mu\nu}=(C_{R,A} D_{A,R} )_{\mu\nu}
\cr
\nonumber\\
& &
=(C_{R,A} B_{A,R} )_{\nu\mu}=(D_{R,A} C_{A,R} )_{\nu\mu}
\cr
\nonumber\\
& &
={1\over 2}({\tilde q_\mu q_\nu\over q^2+2iq_o\epsilon}
+{\tilde q_\mu q_\nu\over q^2-2iq_o\epsilon}).
\end{eqnarray}
\end{document}